# Integration of Heterogeneous Systems as Multi-Agent Systems

*Ammar Lahlouhi*

**Department of computer science, University of Batna, 05000 Batna, Algeria**

ammar.lahlouhi@gmail.com

***Abstract:*** *Systems integration is a difficult matter particularly when its components are varied. The problem becomes even more difficult when such components are heterogeneous such as humans, robots and software systems. Currently, the humans are regarded as users of artificial systems (robots and software systems). This has several disadvantages such as: (1) incoherence of artificial systems exploitation where humans' roles are not clear and (2) vain research of a user's universal model. In this paper, we adopted a cooperative approach where the system's components are regarded as being of the same level and they cooperate for the service of the global system. We concretized such approach by considering humans, robots and software systems as autonomous agents assuming roles in an organization. The latter will be implemented as a multi-agent system developed using a multi-agent development methodology.*

## 1. Introduction

The relation between humans and artificial systems ASs (robots and software systems) is considered in the way that humans are users of artificial systems according to ASs' needs. Such consideration has several disadvantages such as:

- The design of collective employment of ASs by several humans is difficult,
- Such employment can become easily incoherent where the humans roles aren't clear,
- Vain research of users' universal model for all roles and systems.

In this paper, we adopted a cooperative approach in which we consider each system's component as being cooperative and of the same level (peer-to-peer). No component is the server of another but they all serve the global system. We concretized this by considering humans, robots and software systems as autonomous agents assuming roles in an organization. Such an organization will be carried-out by a multi-agent system qualified of heterogeneous society (See Fig. 1). This has several advantages such as:

- It isn't necessary to search a user's universal model. In this approach, each human as a part of the system has a profile that will be defined according to a multi-agent development methodology,
- We use a multi-agent methodology for the development of such societies,
- The system can benefit, in the same time, from the artificial systems capacities but also from the humans' capacities since they are considered as components of the global system.

The heterogeneity of the components (humans, robots and software systems) of such societies poses the problem of their integration solved in this paper by associating specific software systems, qualified of interface agents, for humans and robots. In this paper, we show how we have exploited the cooperation based-approach of interface agents (Lahlouhi, A., et al. 2002), developed initially for humans integration in multi-agent systems, to integrate robots also.

The remaining of the paper is organized in five sections. In section 2, we introduce a classification of the current robotics systems and then we address some of their limitations. Section 3 is a short introduction of the notions of agents and multi-agent systems. We clarify then the notion of heterogeneous societies as multi-agent systems, in section 4, where we explain the relation between heterogeneous societies and multi-agent systems. In section 5, we recall MASA-Method's basic elements, and we explain how we can develop heterogeneous societies in MASA. We illustrate such development by an example of pieces manufacturing society, in section 6. We terminate the paper by a conclusion in section 5, in which we trace some advantages and disadvantages of our approach.

## 2. Current robotics systems

Current robotics systems can be classified in individual robotics systems (implying only a single robot at a time), and Collective robotics systems (including several robots working together). Their limits can be then classified into two types: Individual robotics systems limits regarding the collective robotics, and limits of the robotics systems (individual or collective) not implying humans. The limits of individual robotics are well known for the works on the collective robotics (See for examples (Jung, D., & Zelinsky, A., 1999), (Zlot, R., et al. 2002), (Arai, T., et al.2002)). They can be summarized in the fact that it's impossible for a single robot to make: (1) too difficult tasks, or (2) tasks in parallel on spatially different sites. Furthermore, manufacturing powerful robots (carrying out hard works) is difficult.

The limits of robotics systems not implying humans can be looked in the advantages of some with regard to the others. For examples, the robots excel in the realization of painful, height precision, routine-minded, repetitive, dangerous, intricate ... tasks; while humans excel in the all tasks that humans can make and the robots cannot. Such tasks are numerous. They are those not foreseen in the robots functioning but, especially, those that we cannot endow robots with. The most convincing are, evidently, the intelligent activities. Consequently, robotics systems not implying the humans will be always limited.

A typical situation where an explicit relation among humans and robotics is needed is that of the difficulties that can encounter robots during the achievement of their missions. For example, in some cases, the robot's vision system cannot determine if the object in its front is an obstacle or it is harmless (Fong, T., et al. 2001). In such situations, the only choice a robot has is to continue to execute in bad way or to stop. The humans cannot assist it.

From the relation with humans, we classify the robotics systems in four classes:

- Individual robots completely autonomous (autonomous robots),
- Robots forming completely autonomous society (collective robotics or autonomous robotics society),
- Semi-autonomous robots,
- Collaborative autonomous robotics.

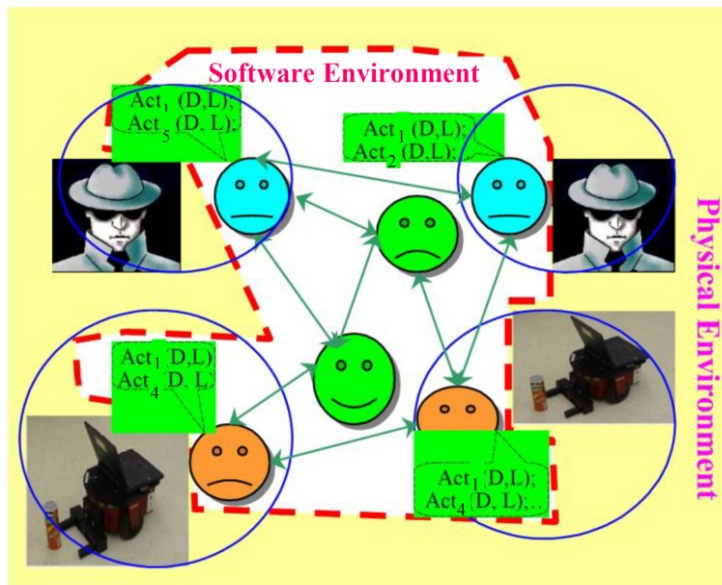

**Comments**

A human (or robot) is considered as an agent, called HR.

To integrate such agents, we associate to the agent HR a soft-ware S, qualified of interface agent. S holds HR's objective. In this figure, S is:

- orange smiley for robots,
- blue smiley for humans.

(green smiley is a software agent)

S holds HR's plan and interacts with HR to carry out the plan's actions, such as [$Act_i$(D, L); $Act_j$(D, L);] where Act is an action on the environment, and D and L are any action parameters.

**Fig. 1: Heterogeneous society**

### 2.1 Autonomous robots

A robot of this category can realize a task in an autonomous way. This task can be a chaining (plan) of more elementary subtasks being able to include sequences and/or loops. The robot is a human's server in this task.

This category can be classified into two subclasses:

1. Supervised robots,
2. Completely autonomous robots.

In the first, the human supplies the tasks plan, which the robot must execute without any help (in an autonomous way). In the second, the human establishes high level goals, which the robot will accomplish in an independent way by elaborating a plan itself. The difference between the two classes is in the nature of the goal (Fong, T., et al. 2001). In the first case, the goals are limited and the human realizes the tasks planning. In the second case, the goals are more abstract and the robot is responsible for the plan's elaboration. Let us indicate that the robot can be completely autonomous only for some simple applications (Fong, T., et al. 2001).

The autonomous robots are interesting for the numerous applications which they have in the daily and professional life but, furthermore, because they constitute the base of all the other categories. The limits of this category are those of the individual robotics and those of robotics systems not implying directly the humans.

## 2.2 Robotics autonomous societies

Robots organized in an autonomous society can make a common global task, autonomously. In this case, the autonomy concerns not only the robots but also all what is internal to the society. Robots are then human's servers in this task. The study of collective robotics extends naturally that on the individual robots systems, but it is also an independent discipline. This class can be also subdivided into two subclasses:

1. Supervised societies,
3. Completely autonomous societies.

Most of the proposed architectures for the collective robotics, such as AuRA (Arkin, R. C., & Balch, T., 1997) and ALLIANCE (Parker, L. E., 1998), concentrate on the intensification of the robot teams' autonomous capacities rather than on their operation with humans.

The research that incorporated the human operator (such as ROBODIS (Surmann, H., & Theißinger, M., 1999), RAVE (Dixon, K., et al. 1999), MokSAF (Payne, T. R., et al. 2000) and MissionLab (Arkin, R. C., et al. 1999)) consider the human as being, especially, available for the initial planning. The collective robotics advantages are those remedying the individual robotics limits and its limits are those of robotics systems not implying the humans.

## 2.3 Semiautonomous robots

A semiautonomous robot can realize several elementary tasks in an autonomous way, but only one task at a time (The chaining plan of such tasks is missing in such robots). Such robot serves the human by realizing, every time, one task. These robots can be considered as specialized tools, which are exploited by the humans in various ways. Robots are then considered as human servers who dictate them, every time, what they have to make. The human possesses then the robot's command. In this case, we have two roles:

- That of a tool knowing and able to make elementary things in a completely automatic way,
- That of a human who has to exploit this knowledge and this capacity to reach various objectives.

The importance of such robots lies in their use in situations where the human's supervision is very sought such as surgery. However, the difficulties of this category are, in part, those of the individual robotics. Furthermore, such robots are limited to the formed persons (experts' users) because of considerable difficulties and risks they put. Because all the command decisions are human-dependent, the performance of the entire system is then bounded by the human skills. Several human factors, including skill and training, influence the system's functioning. Other factors, such as the communication's band width, in the case of the tele-operation, can also influence the operational efficiency.

If a semiautonomous robot has a difficulty, it is up to the human to get it out from this situation by indicating to the robot what to make. This can be qualified as an advantage, with regard to the autonomous robots. However, some situations are hard to solve, such as where the human does not:

- Know that the robot has a difficulty, such as a situation of tele-operation, or

- The difficulty's nature which has the robot; a robot remaining immovable and its user does not know for what reason.

In such situations, the only choices, which the robot has, are to continue to execute badly or to stop. Finally, the human is completely mobilized to command the robot. Such mobilization would overload the human being and/or sub exploits the robot (Fong, T., et al. 2001).

### 2.4 Collaborative autonomous robots

The robots of this category can realize tasks autonomously while having the capacity to ask the human to realize given tasks on behalf of the robot. The human is considered as being a source (Fong, T., et al. 2001) for the robot (supply information, resolves some ambiguities, etc.). The robot serves the human for a pre-established task while asking the human to realize particular tasks.

The most representative of these works is the Fong's collaborative approach (Fong, T., et al. 2001). In such approach, the human must be at the robot's disposal to serve it according to its needs.

The advantage of the robots of this category, with regard to the previous categories, is that they allow humans to compensate the autonomy's inadequacy in some situations while decreasing the human's overload and increasing the robot's exploitation. Among its disadvantages, we can cite those of the individual robotics. Furthermore, such systems cannot beneficiate completely from humans' capacities since humans are solicited only to realize tasks in some situations and they are not involved completely in the global system.

### 2.5 Conclusion

The major difficulties of current robotics systems reside in:

- the inability of isolated robots to accomplish some hardest and spatially distributed tasks,
- the limits of the robots autonomy where they always need humans.

Such systems fail also in finding equilibrium between the tasks that must be accomplished by humans, and those which must be accomplished by the robots. Finally, they fail in their visions of the software systems where such systems are considered as robots functioning only.

In the following sections, we explain the cooperative approach to the development of heterogeneous societies', which use both human faculties and software systems. The development of such societies was done using a multi-agent development methodology.

## 3. Agents and multi-agents systems

In order to clarify the reasoning followed in the development of heterogeneous societies, we introduce some notions about the vision of humans as systems and the cooperation between heterogeneous systems, including humans, robots and software systems. In subsection 3.1, we explain the systems properties. We show, particularly, the relationship between the general notion of system with software systems, robots and humans. We explain, then, the agent's notion as a system, in subsection 3.2, to introduce cooperation aspects in multi-agent systems, in subsection 3.3.

### 3.1 System's properties

A reactive system is the one that perceives its environment by means of its sensors and acts on it by means of its effectors. A communicative system uses some of its sensors to receive messages from other systems and uses some of its effectors to submit messages to others according to a pre-established communication protocols. Among the system's aspects two are fundamental: control and functioning. The system's control commands the sensors to perceive the environment and the effectors to act on such environment. It acts on its sensors and effectors by conforming to a system's functioning.

A classical system (CS) is considered as a server used according to a functioning and pre-established communication protocols. Its functioning is reduced to knowledge procedures. The user (an agent who is, usually, a human) invokes the system's knowledge procedures according to a plan that he (she) possesses that CS does not know. The role of CS's control is limited to interpreting the communication's protocols and the knowledge, and to command sensors and effectors according to external systems demands.

### 3.1.1 System's autonomy (self-control and pro-activity)

The system's autonomy can be seen from several viewpoints: energy, resources, decision-making, etc. From the control's viewpoint, CS can be considered as being autonomous. This means that the system, without any external intervention, commands its sensors and its effectors and interprets its knowledge and its communication protocols. However, a CS isn't functionally autonomous in a manner that it depends on another system, which indicates every time what it must make.

A functionally autonomous system can take the initiative to act on the environment; the system is called proactive. We obtain a proactive system by endowing the system with an objective allowing its control to make decisions on what it has to make. The system will be then directed by such an objective, which is proper to it, not by external interventions. The system rejects then external interventions that are not compatible with its objective.

To help a proactive system to make decisions on external interventions (asking it to accomplish actions) it is necessary that these demands must be accompanied with elements others than asked action, in contrast to the classical systems. These elements concern, in particular, a performative clarifying the demand's nature (wish, order, etc.). This implies that the communication protocols of proactive systems must be based on speech acts. Consequently, the proactive system's functioning consists in two parts: knowledge and system's objective supplied with speech acts protocols.

### 3.1.2 System's social ability.

A system, even though it is autonomous, lives in an environment with other systems with which it cooperates in a direct or indirect way. This cooperation is indirect if the system acts on the environment without coordinating (communicating and synchronizing) its actions with other systems. It's direct in the case of system's actions are coordinated with other systems. In this case, the system is qualified as being with a social ability.

## 3.2 Agents

Several works were used agent technology (such as (Mellah, H., 2011)) with different meanings. In this paper, we take the vision considering an agent as a system. In such vision, see Jennings (2001), an agent is considered as a system enjoying the properties of autonomy, reactivity, proactivity and social ability. An agent must be then endowed with:

- A control, which can be procedural, declarative or intelligent,
- Sensors and effectors which we subdivide into two classes:
    - Class 1. those used for the communication with other agents,
    - Class 2. those used to perceive and to act on the environment, such as the wheels of a robot,
- Communication protocols, which must be based on speech acts. Their descriptions are functions of perceptions and actions on the environment by means of the agent's sensors and effectors of the first class. They can be considered as particular knowledge,
- Knowledge described according to perceptions and actions on the environment by means of sensors and effectors of the second class,
- An objective described according to the agent's knowledge procedures and its communications. These last ones allow invoking the knowledge's procedures of other agents.

The agent's control is responsible of:
- Interpretation of the agent's objective, its communication protocols and its knowledge's procedures,
- Command its sensors and effectors.

## 3.3 Multi-agent systems

A multi-agent system is composed of agents assuming roles in a given organization. It possesses the system's fundamental aspects, which are: control and functioning. The agents are autonomous and pro-active systems. Consequently, the control and functioning of a multi-agent system are distributed on its agents. The only direct relation that can exist between the agents is that of their communication.

The agents are cooperative. We consider the cooperation as being an achievement of a common global task. Since the agents are autonomous, it is necessary to distribute this global task on the agents. Each agent will have its own subtask (its individual task) of the organization's global task. The achievement, in a coordinated way, of the agents' individual tasks must allow reaching the organization's global objective.

## 4. Heterogeneous societies as multi-agent systems

Heterogeneous societies' components are humans, robots and software systems. If the consideration of software systems and robots as systems is very common, this is less obvious for the humans. In this section, we try to explain the relation between such components and the agents (in subsection 4.1), we recall the cooperation based approach (in subsection 4.2), and the integration of heterogeneous agents so they constitute a multi-agent system (in subsection 4.3).

### 4.1 Heterogeneous societies components as agents

#### 4.1.1 Humans

Humans are agents by Excellency. The agent's scientific notion itself is borrowed from that of the human. In theory, a human doesn't make any task without being aware of its meaning. He has the properties of control's autonomy, reactivity, pro-activity and social ability. However, the human's objective can be not compatible with that of the global system where we would place him. To overcome this difficulty, we require from a human to temporary change its implicit objective to follow a pre-established plan so he can be accepted as a part of a given system. This isn't new in the humans' societies or its relation with artificial systems. For example, to integrate him in a given work, we demand to a human to behave in a special way. Another example is that of using a tool where the human will follow the constructor's instructions. There is then always a required preliminary condition so a human can exploit a classical system. However, in contrast to this, in the cooperation-based approach, the human is directed by a clear objective.

#### 4.1.2 Robots

A robot is an attempt to reproduce the human. The robots developers try to endow them with the human's characteristics. It has then several agents' characteristics. However, the developers' objective, which is that of human's reproduction, hasn't been reached completely. Consequently, the robots don't have some of the agents' characteristics. A robot is a physical system and has its own control. When we submit it under another system's supervision, such as a computer system or a human, it executes the commands dictated by such supervisor system. In this case, the commands sequence submitted to the robot constitute its functioning, not its control. A simple robot has then reactivity and control's autonomy characteristics but it isn't proactive. When we endow a robot by a simple explicit objective, such as that affected to special tasks or that with functioning, it becomes pro-active. But, to endow them with a social ability, these robots must have the ability of communication with other systems.

#### 4.1.3 Software system

Before we associate it to a processor, the software system has not a control's autonomy. It is used by a hardware system as a functioning (a plan) to direct it in accomplishing some tasks. Consequently, the software system is considered as a functioning. When we associate a processor and peripherals to software, we obtain a system that has a control's autonomy and reactivity. Implicitly, what we consider as software system is the result of such association. In practice, we associate to a software system, which we will consider as agent, an independent thread representing its virtual control. We can then qualify a software agent as a virtual agent.

The reactivity of a software system is obtained by the command of its peripherals that constitutes its sensors and effectors. However, a classical software system isn't proactive and hasn't a social ability. It's then necessary to endow it with such characteristics.

Classical software systems have, in reality, an implicit objective that is the one of serving its users. It waits for the user's demand and reacts to it according to its knowledge. For example, the object-oriented system's control proceeds as follows:
1. Creates some system's objects, and displays references to some objects, in form of icons. The latter constitute an object-oriented interface,

2. Waits that a user selects (from a contextual menu, for example) one of these icons (O) and one of its actions or methods (M),
3. Executes M on O's state, and go to 2.

Such objective doesn't allow the system to make initiative in executing its methods without being demanded by another part. To make a classical system pro-active, we must change its implicit objective by an explicit one allowing it to take initiative. The means for this depends on the software classical system's model.

### 4.2 Cooperation-based approach for interface agents development

Abstractly, in the cooperation-based approach, we consider a human (or a robot) as an agent HR that assumes one or more roles. Conceptually, we associate to HR a software system S that holds the HR's objective. S is a software agent qualified of interface agent.

The interface agent for a human (hS) has neither the human's intellectual abilities nor its resources exploitation. We supplement this lack by an interaction between hS and a human. The agent HR consists then of hS completed by the human's intellectual abilities and his resources exploitation. Similarly, the interface agent for a robot (rS) hasn't the robot's physical abilities concerning the actions achievement on the physical environment. The agent HR will then be consisted of rS completed by a robot's physical abilities. Since HR is an agent belonging to a multi-agent system, the derivation of the required characteristics of HR and then those of rS or hS will follow a multi-agent development methodology.

In the cooperation-based approach, the actions' plan is qualified of individual task and must be explicit. Its base is the roles that the robot or the human must assume in a given organization (see section 5). It is then determined during the multi-agent system (MAS) design so that the MAS's functioning will be coherent. It is the interface agent's responsibility to hold the individual task, and to ensure its execution. In the case of the human, this execution consists of:

- Sequencing actions plan,
- Requesting from the human to execute his knowledge's procedures,
- Requesting from the software agents (including other interface agents) to execute their actions.

In the case of the robot, it consists of:

- Sequencing actions plan,
- Executing the knowledge's procedures by communicating to the robot the necessary commands,
- Demanding to software agents (including other interface agents) to execute their actions.

The difference between a robot's interface agent and that of a human is in the communicated objectives level. The human's interface agent can communicate with the human in high level objectives. This is not the case with a robot where we can communicate it only elementary actions in a specific command language. The difficulty is that in the design level where it isn't convenient to manipulate such commands, directly. Consequently, we use, at the design stage, high level objectives that we must implement in the specific command language, at the implementation stage. This means that the human's interface agent holds only the objective and the human possesses the knowledge's procedures. However, the robot's interface agent must hold also the knowledge's procedures that are functions of the robot's commands.

### 4.3 Heterogeneous agents integration as multi-agent system

As we explained it, software systems, robots and humans can be considered as classical systems that we can endow them with the agents properties, which are those of control autonomy, reactivity, pro-activity and social ability. The difficult task is that of the heterogeneous agents' integration so that they can cooperate harmoniously. We overcome this difficulty by exploiting the cooperation-based approach for the interface agents (Lahlouhi, A., et al. 2002). One of the interface agent's purposes is to establish a sound communication relation between the virtual (software) environment and the non-software systems, and vice versa. Furthermore, such interface agents are software systems that can communicate with other software agents including other interface agents. As a consequence, we can establish a communication among software and non-software agents and among non-software agents themselves by means of interface agents.

Being given that the communication's relations are the only relation that exists between the agents, the result is then a multi- agent system with heterogeneous agents. The communication's relations are very useful between robots and humans. They are also useful for the communication among humans who cannot communicate directly (situated in remote spaces). But, sometimes, these relations are even useful for humans situated nearly (among video game's players, in the case of the communication of voluminous information, etc.). However, in many situations, the humans communicate directly among them in a suitable way. It is then the responsibility of the system's developer to choose the most suitable way to communicate between two humans who are a part of a heterogeneous society.

In heterogeneous society, we can charge the software agents with:
- Executing tasks concerning the data processing automation: calculations, storage, research...,
- Making some decisions.

We can then charge the robots by achieving:
- Actions on physical environment, such as object transportation for mobile robots, manufacturing for tool machines (decided by themselves, by humans and/or by software agents),
- Information and events perception such as the axes positions, scene (camera), etc.

Whereas, the human can be charged by:
- Achieving treatments of intelligent tasks,
- Making some decisions,
- Achieving actions on the physical environment decided by them, by software agents and/or by robots.

## 5. Heterogeneous societies in MASA

In this section, we show how we can develop heterogeneous societies in MASA project. In subsection 5.1, we present briefly MASA-Method (see (Lahlouhi, A., 2006), for more details on MASA-Method's description). In subsection 5.2, we explain how we derive interface agents in MASA.

### 5.1 MASA-Method

MASA project (Multi-Agent modeling of Systems based on Autonomous and heterogeneous components) comprises three parts:
- MASA-meta-model which is a multi-agent meta-model,
- MASA-Method which is an organizational, complete and mixed multi-agent methodology (see (Lahlouhi, A, & Sahnoun, Z., 2002), for a classification of multi-agent methodologies),
- MASA-applications which is a set of multi-agent applications.

These applications, which are also significant ones, have the objective of correcting, enriching and improving MASA-meta-model and MASA-Method. They concern, mainly:
- Software development process (Lahlouhi, A., 2006),
- Networks exploration by mobile agents (Lahlouhi. A., 2014),
- Image processing (Melkemi, K. E., et al. 2006),
- Heterogeneous societies, in this paper.

#### 5.1.1 Organizations in MASA

The organization's structure is composed of roles, and communication relations between them. In reality, we create an organization to achieve a given goal. From the methodological viewpoint, the developer puts in his mind an objective, and he creates, then, an organization to achieve such goal. Obviously, the objective will be merged in that of the organization once it is established. However, methodologically, it exists. Consequently, we define an organization as a pair (Organization's structure, Global objective).

We can replace a declarative expression of objective by a procedural expression "achieve goal". The task "achieve goal" is the organization's global task. Consequently, we define an organization as a pair (organization's structure, global task).

In MASA, we employ colored Petri nets (Jensen, K., 1994) (CPN) for describing global tasks. We made the global task's description as follows:

- A transition contains a role attached to one of its knowledge's procedures,
- Any entry arc to a transition is labeled with a procedure's input,
- Any exit arc from a transition is labeled with a procedure's output,
- The places are means of communication between such procedures (outputs for some procedures and inputs for others).

### 5.1.2 Agents and multi-agent systems description in MASA

A multi-agent system is composed of agents assuming roles of a given organization. The agents are cooperative in a way that their objectives are derived from those of the organization according to a roles attribution. Since the organizations' global objectives in MASA are described using CPN, the agents' objectives will be also expressed using CPN.

In ordinary Petri nets, the communication isn't explicit. This makes difficult to express the communication between agents in individual tasks. Research works on the agents' behaviors modeling employs autonomous Petri nets where they represent communications by arcs (see (Xu, H., & Shatz, S. M. 2001), for example). This does not express clearly the agents' autonomy, which is a fundamental characteristic. For the description of individual tasks, we employ then non-autonomous colored Petri nets (see (David, R., & Alla, H., 2010)). In autonomous Petri nets, the transitions are subjected to internal Petri nets conditions only, in particular, places marking. In non-autonomous Petri nets, the transitions are subjected to external events also.

In MASA, we perceive communication messages as events attached to transitions. A transition's outgoing arc schematizes the message's emission while an entering arc schematizes a message's reception. We do not consider what occurs between the message's emission and its reception. This belongs to implementations of the communication protocols.

### 5.1.3 MASA-Method's methodological process

The MASA-Method's methodological process consists in:

- Creating an organization satisfying the system's requirements. This includes, mainly, the description of its communication structure and its global task (GT),
- Derivation of the multi-agent model satisfying such organization. This includes, mainly, the attribution of the roles to agents and the derivation of the agents' individual tasks from GT,
- Multi-agent model's implementation as a distributed object oriented multi-system.

## 5.2 Interface Agents derivation in MASA

The necessary characteristics of a human and/or a robot, which is a system's component, will be known by following a multi-agent methodology MASA-Method. However, the characteristics that we must attribute to an interface agent are derived as follows:

- **Environment's sensors and effectors:** they are means of interaction between the interface agent and the human (respectively, robot), such as keyboard and screen for the human (resp. interfaces for the robot). The interface agent uses them not to perceive the environment and to act on it, directly, but especially to be informed by the human (resp. by the robot) of the environment's state of the human (resp. robot) and to demand to the human (resp. to the robot) to act on such environment,
- **Communication's sensors and effectors:** similar to those of communication between software agents,
- **Communication protocols:** the same as those used by the software agents,
- **Knowledge's procedures:** See below,
- **Individual task:** we derive the individual task from the CPN's description of the global task as for other agents,
- **Control:** similar to that of software agents.

Knowledge's procedure of an interface agent for a human presents him the procedure's name and its data and/or receives from him its execution's results. If the communication means with the human are the usual ones (keyboard, screen, mouse...), displaying data and/or inputting results can be complex. They can be formatted texts, graphs and/or images. In this case, the associated procedures must use tools enabling them to display such data and/or to assist the human to enter results. A knowledge procedure of an interface agent will be then described as follows:

- Display the procedure's name and a description of it,
- Display its data if there is any,
- Recover results if there is any.

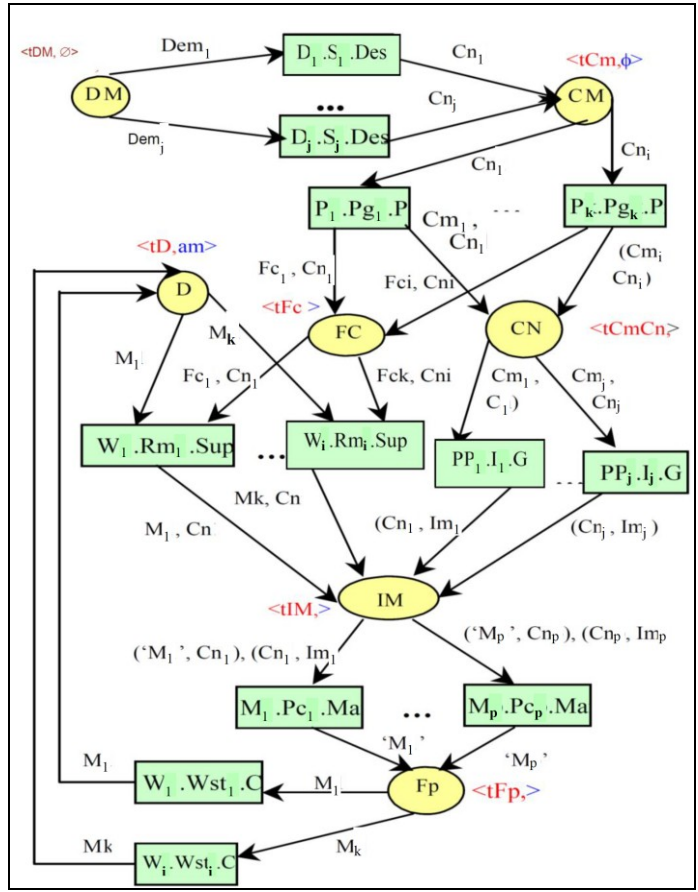

| **Summary of the CPN** | | |
| --- | --- | --- |
| (Transition = Role.Object.Operation) | | |
| Role | Object | Operation |
| $D_i$: designer | $S_i$: schema | Des: Design |
| $P_i$: Programmer | $Pg_i$: program for manufacturing | P: program |
| $PP_i$: Translator (from program to memory image) | $I_i$: memory image associated to $Pg_i$ for manufacturing | G: generate memory image associated to a program |
| $W_i$: Worker | $Rm_i$: Raw material | Sup: Supply Machine with |
| $M_i$: Machine | $Pc_i$: piece | Ma: Manufacture |
|  | $Wst_i$: wast | C: Clear machine |

**Fig. 2: Case study - Global task of the organization of pieces manufacturing.**

## 6. Case study: Pieces manufacturing society

The system of pieces' manufacturing takes the customer's demands Dem. Then, the designers create a schema (design) Cn satisfying each demand. Such schema will be transformed into a manufacturing program Pg by skilled programmers. Pg will be then translated into a memory image Im by a software translator PP. The worker W then supplies the digital machine with raw material Rm. The machine uses the memory image and the raw material to manufacture pieces. Once manufacturing is done, the worker cleans C the machine and removes waste Wst. Such scenario is described in the CPN of Fig. 2.

**Tab. 1: Case study - Organization's description**

| Identifier | PMO | | | |
|---|---|---|---|---|
| Description | Organization for pieces manufacturing | | | |
| Task | Fig. 2 | | | |
| Roles | Identifiers | Model | Identifiers | Model |
| | $D_1 … D_n$ | DM | $W_1 … W_i$ | WM |
| | $P_1 … P_k$ | PM | $M_1 … M_p$ | MM |
| | $PP_1 … PP_j$ | PPM | | |
| Communication relation | Analytic | | | Graphic |
| | Role 1 | Role 2 | Role 1 | Role 2 |
| | $M_i$ | $W_i$ | $P_i$ | $PP_i$ |
| | $M_i$ | $PP_i$ | $D_i$ | $P_i$ | Fig. 3.a |
| | $P_i$ | $W_i$ | | |
| Legend | Symbol | Description | Symbol | Description |
| | D | Designer of model DM | W | Worker of model WM |
| | P | Programmer of model PM | | |
| | PP | Memory image generator of model PPM | M | Machine of model MM |

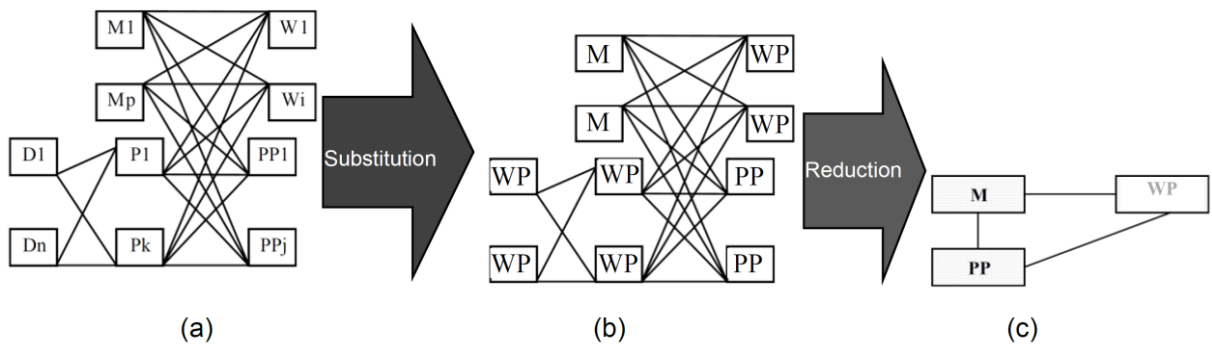

**Fig. 3: Case study - Communication relation between roles, substitution of the roles and creation of communication links between agents.**

### 6.1 Development of the multi-agent system for pieces manufacturing

In the following, we present a model of multi-agent system according to the following process (see (Lahlouhi, A., 2006) for further details on MASA-Method's development processes):

1. Develop an organization's model (see Tab. 1): This concerns mainly the development of the organization's task (Fig. 2) and the communication relations between roles (Fig. 3.a),

2. Derivation of a multi-agent model (Tab. 2): This step includes identifying agents and assigning such agents to roles. In our case study, we identified three agents:
    a. WP: Worker-Programmer (Human),
    b. PP: Translator (Software system),
    c. M: Machine (Robot).
3. Derivation of the communications between the agents: In the organization's communication structure (Fig. 3.a), we substitute the roles by the agents assigned to them (Fig. 3.b) to lead to a simplified model of communication between agents (Fig. 3.c),
4. Derivation of the multi-agent's global task (Fig. 4). The latter is the same as that of the organization's task in Fig. 2; in which we substitute the roles by the agents assuming them,
5. Simplification of the multi-agent global task; if it can be simplified as indicated in Fig. 5. In the latter, we regrouped identical transitions into one transition. The parts of the agents' tasks are identified based on the transitions including the agents' identifiers illustrated by coloring them,
6. Derivation of the agents' models (Fig. 6): we give such models in related tables as follows (further details on such derivation are given in subsection 6.2): Agent M (Tab. 3), Agent WP (Tab. 4), and Agent PP: we don't give the table describing PP. It is obvious.
7. Implementation of agents models which constitute an implementation of the multi-agent model (not given in this paper). Such implementation will be based on the tables (Tab. 3 and Tab.4) describing the agents.

It is important to note that such process doesn't make a distinction between software, human and robot agents. It is at the implementation phase that we distinguish between such agents particularly in the implementation of their knowledge procedures. For the humans, such procedures ask the human agent to carry out some actions where for the robots it commands them to do things.

## 6.2  Derivation of the agents tasks

To derive the agents' tasks (their associated CPN) from that of the multi-agent's global task, we proceed as follows. Firstly, we identify the parts of the agents' tasks as shown in Fig. 5. Secondly, we extract the CPNs of the different agents while adding the communication between the agents (see Fig. 6); i.e., sending messages from senders and receiving such messages by receivers (see (Lahlouhi, A, 2006) for further details on methodological guiding). Note that the agents' identifiers are needless in the transitions of the agents' CPN. The messages are based on speech acts (Smith, I. A., & Cohen, P. R., 1996) and they are composed as follows; as indicated in Fig. 6:

1. (Receiving agent, performative, action, (action parameters)), for sending a message. It is represented by a place containing the action's parameters and a transition labelled with "the agent's identifier followed with S (for Sensor)" which is followed by an action "Act Emission, AE". For example, for a transition labeled "WPS.AE", WPS indicates a communication sensor of the receiver WP. Such a sensor is an object of WP, in an object-oriented terminology. The same sensor will be used at receiving the message. Finally, an arrow indicates that the message exits such transition,
2. (Sent agent, performative, action, (action parameters)), for receiving a message. It is represented by a transition labelled with "the agent's identifier followed with S (for Sensor)" followed by a place containing the action's parameters. An arrow showing the message entering such transition as an event, which indicates that such CPN is not autonomous, i.e., firing such transition is conditioned by the reception of such message.

**Tab. 2: Case study - Multi-agent system model**

| Identifier | PM | | |
|---|---|---|---|
| Global task | Fig. 4 | | |
| Description | Imitates the PMO's organization | | |
| Organization | PMO | | |
| Roles attribution (see Tab. 2) | **Agent identifier** | **Roles** | |
| | WP | $P_1, …, P_k, W_1, …, W_i, D_1, …, D_j$ | |
| | M | $M_1, …, M_p$ | |
| | PP | $PP_1, …, PP_i$ | |
| Communication links | **Graphic** | **Analytic** | |
| | | **Agent 1** | **Agent 2** |
| | Fig. 3.c | WP | M |
| | | PP | M |
| | | PP | WP |
| Legend (For roles see Tab. 1) | **Acronym** | **Description** | |
| | **WP** | Programmer-Worker | |
| | **M** | Machine | |
| | **PP** | Memory image generator | |

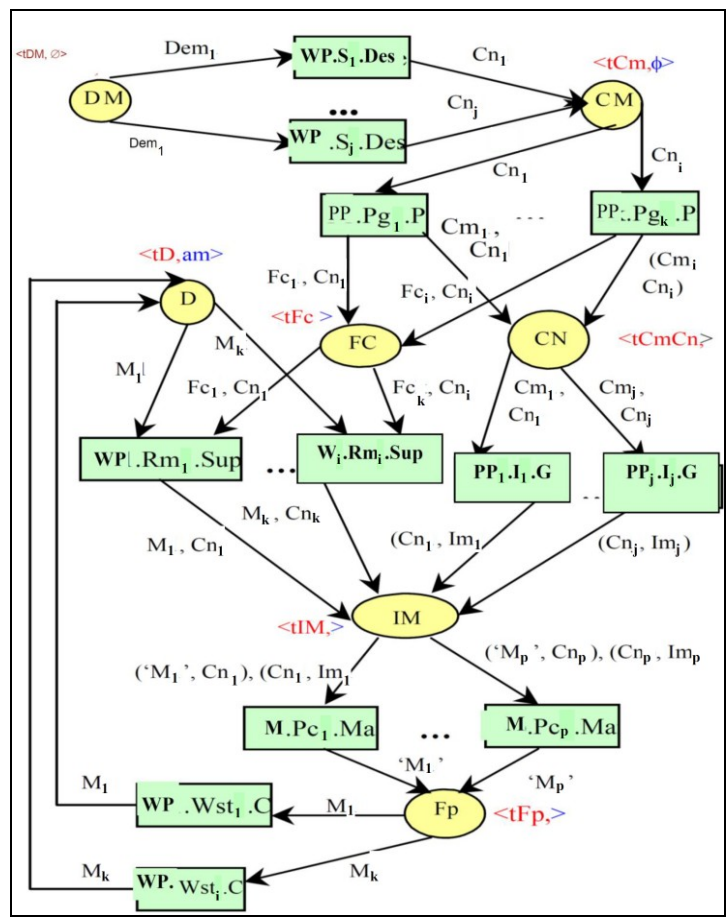

**Fig. 4: Case study – MAS's Global task for pieces manufacturing**

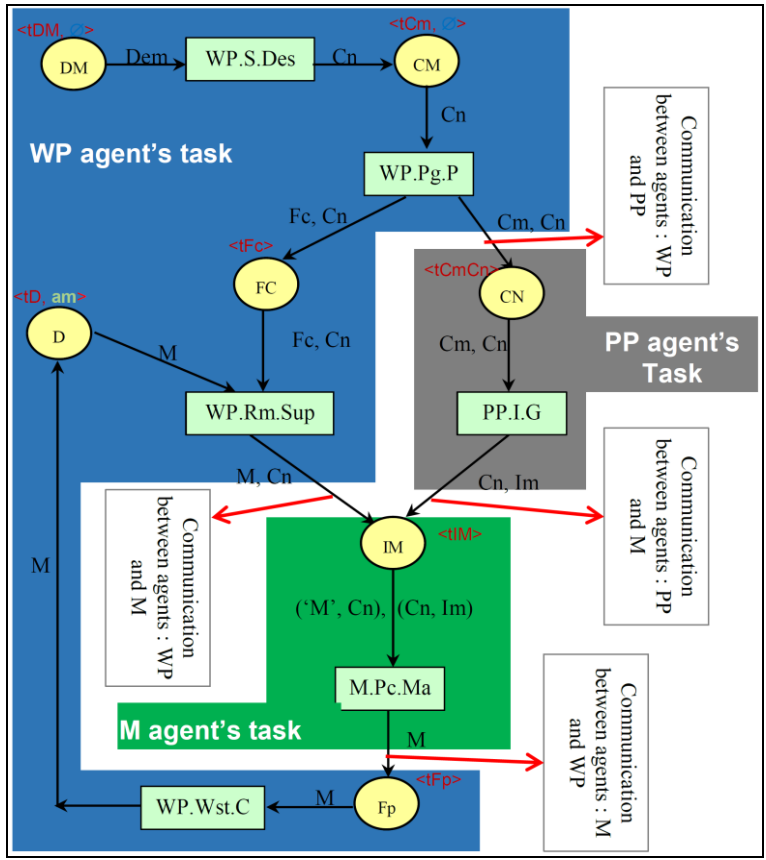

**Fig. 5: Case study – MAS's Global task simplified**

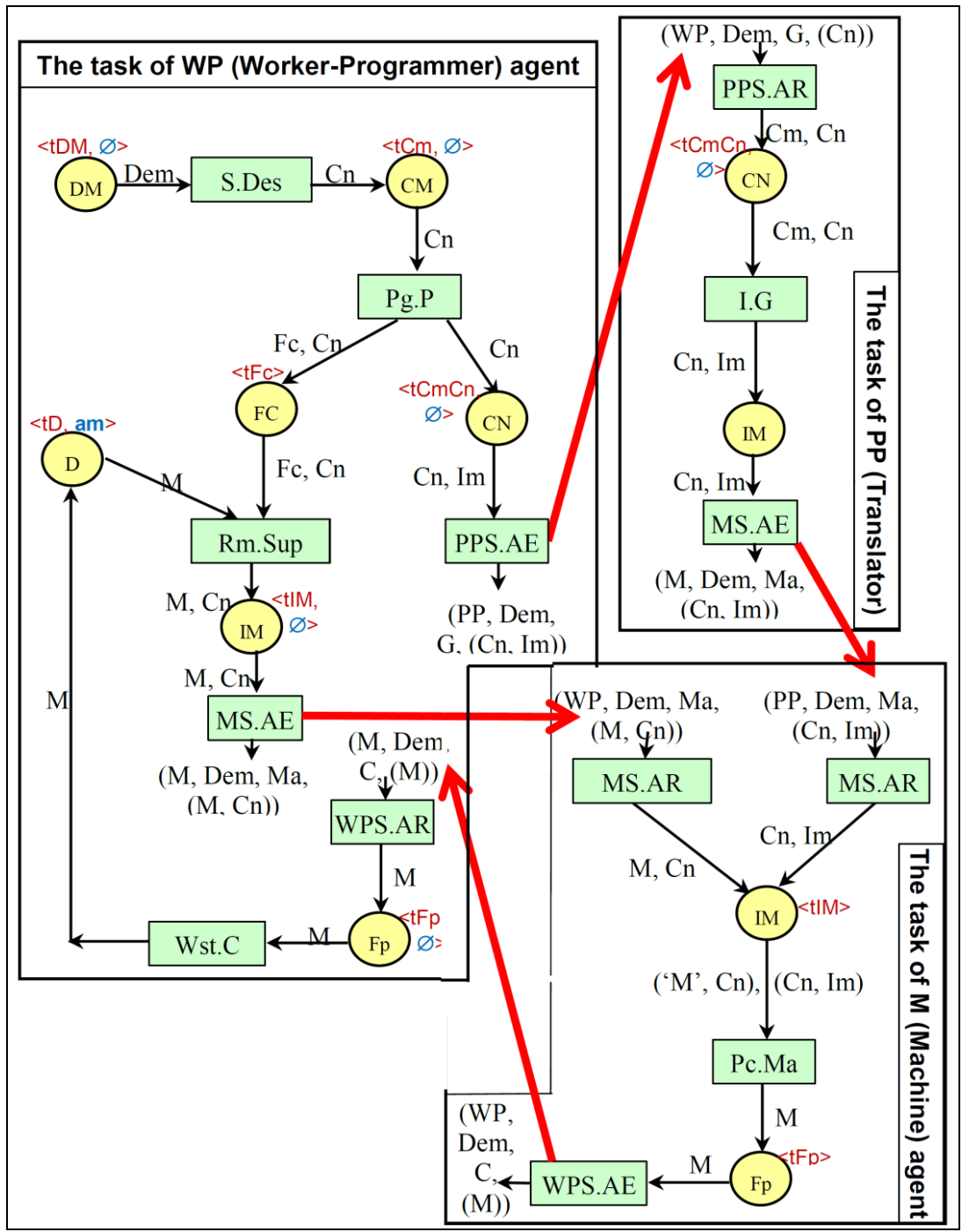

**Fig. 6: Case study – The agents' tasks resulted from the decomposition of the MAS's global task after adding the communication (showed by red arrows) transitions between the agents**

**Tab. 3: Case study - Agent model description – M (Machine agent)**

| Identifier | | M (Machine agent) | | |
|---|---|---|---|---|
| Multi-agent model | | PM | | |
| Control | | Procedural | | |
| Task description | | See Fig. 6 | | |
| Communications | Sensors | //for receiving messages<br>MS: …; | | |
| | Protocols | // Describing protocols for sending and<br>// receiving messages<br>MS.AR (S, P, A, Par): ...; | | |
| Environment relation | Objects | Pg, I, Rm, Pc | | |
| | Knowledge | Pc.Man: Manufacture a piece using the memory image and the raw material | | |
| | Sensors | WPE: ..., … | | |
| | Effectors | WPR: ..., … | | |
| Legend (For roles see Tab. 1) | Acronym | Significance | Acronym | Significance |
| | P | Performative | A | Action |
| | R | Receiver | Par | Parameters |
| | S | Sender | Pc | Piece |
| | Man | Manufacture | Other symb. | See Tab. 4 |

**Tab. 4: Case study - Agent model description – WP (Worker-programmer agent)**

| Identifier | | WP (Worker-programmer agent) | | |
|---|---|---|---|---|
| Multi-agent model | | PM | | |
| Control | | Intelligent | | |
| Task description | | See Fig. 6 | | |
| Communications | Sensors | //for receiving messages<br>WPS: …; | | |
| | Protocols | // Describing protocols for sending<br>// and receiving messages<br>WPS.AE(S, P, A, Par): ...; | | |
| Environment relation | Objects | S, Pg, Rm, Wst | | |
| | Knowledge | Pg.P: produce the program for manufacturing the design | | |
| | Sensors | PPE : ...; … | | |
| | Effectors | AMR: ...; … | | |
| Legend (For the roles see Tab. 1) | Acronym | Significance | Acronym | Significance |
| | S | schema | Wst | waste |
| | Pg | program for manufacturing | WPS | WP sensor |
| | Rm | Raw material | Other symbols | See Tab. 3 |

## 7. Conclusion

In this paper, we described MASA's approach for the development of heterogeneous societies as multi-agent systems. MASA's approach consists in integrating heterogeneous agents (humans, robots and software systems). Such integration is made by using special software systems, called interface agents. The derivation of the interface agents' characteristics is made by using MASA-Method, a multi-agent development methodology. The tasks of the derived agents are synchronized using Colored Petri Nets. We showed the application of the approach by a case-study of manufacturing pieces that uses heterogeneous agents: software, robot and human.

The MASA's development approach of heterogeneous societies inherits the advantages of both collaborative and collective robotics approaches. The inheritance of the advantages of collaborative approach resides in the fact that the resulted robotics system contains humans where the robots can ask them for services. The inheritance of the collective robotics advantages resides in the fact that heterogeneous society can contain several robots, which cooperate to realize a common global task. In addition, MASA's approach has several other advantages, with regard to previous approaches, such as the following:

- Heterogeneous agents (software, robots and humans) can share tasks in a coherent way,
- Using multi-agents methodology for the characteristics' derivation of the agents (robots and humans) and, consequently, for those of their interface agents,
- The same system can benefit from the (intellectual and/or physical) abilities of humans and robots,
- The cooperation can be managed easily between humans, between robots and/or between robots and humans through a computer network.

**JEL Classification: M15**